\documentclass[14pt,twocolumn]{aastex62}
\usepackage{lineno}

\graphicspath{{./}{figures/}}

\submitjournal{ApJ}

\shorttitle{Search for Neutrino Emission from Pulsar Wind Nebulae}
\shortauthors{IceCube Collaboration}

\begin{document}
\email{analysis@icecube.wisc.edu}
\title{IceCube Search for High-Energy Neutrino Emission from TeV Pulsar Wind Nebulae}

 \affiliation{III. Physikalisches Institut, RWTH Aachen University, D-52056 Aachen, Germany} \affiliation{Department of Physics, University of Adelaide, Adelaide, 5005, Australia} \affiliation{Dept. of Physics and Astronomy, University of Alaska Anchorage, 3211 Providence Dr., Anchorage, AK 99508, USA} \affiliation{Dept. of Physics, University of Texas at Arlington, 502 Yates St., Science Hall Rm 108, Box 19059, Arlington, TX 76019, USA} \affiliation{CTSPS, Clark-Atlanta University, Atlanta, GA 30314, USA} \affiliation{School of Physics and Center for Relativistic Astrophysics, Georgia Institute of Technology, Atlanta, GA 30332, USA} \affiliation{Dept. of Physics, Southern University, Baton Rouge, LA 70813, USA} \affiliation{Dept. of Physics, University of California, Berkeley, CA 94720, USA} \affiliation{Lawrence Berkeley National Laboratory, Berkeley, CA 94720, USA} \affiliation{Institut f{\"u}r Physik, Humboldt-Universit{\"a}t zu Berlin, D-12489 Berlin, Germany} \affiliation{Fakult{\"a}t f{\"u}r Physik {\&} Astronomie, Ruhr-Universit{\"a}t Bochum, D-44780 Bochum, Germany} \affiliation{Universit{\'e} Libre de Bruxelles, Science Faculty CP230, B-1050 Brussels, Belgium} \affiliation{Vrije Universiteit Brussel (VUB), Dienst ELEM, B-1050 Brussels, Belgium} \affiliation{Dept. of Physics, Massachusetts Institute of Technology, Cambridge, MA 02139, USA} \affiliation{Dept. of Physics and Institute for Global Prominent Research, Chiba University, Chiba 263-8522, Japan} \affiliation{Dept. of Physics and Astronomy, University of Canterbury, Private Bag 4800, Christchurch, New Zealand} \affiliation{Dept. of Physics, University of Maryland, College Park, MD 20742, USA} \affiliation{Dept. of Astronomy, Ohio State University, Columbus, OH 43210, USA} \affiliation{Dept. of Physics and Center for Cosmology and Astro-Particle Physics, Ohio State University, Columbus, OH 43210, USA} \affiliation{Niels Bohr Institute, University of Copenhagen, DK-2100 Copenhagen, Denmark} \affiliation{Dept. of Physics, TU Dortmund University, D-44221 Dortmund, Germany} \affiliation{Dept. of Physics and Astronomy, Michigan State University, East Lansing, MI 48824, USA} \affiliation{Dept. of Physics, University of Alberta, Edmonton, Alberta, Canada T6G 2E1} \affiliation{Erlangen Centre for Astroparticle Physics, Friedrich-Alexander-Universit{\"a}t Erlangen-N{\"u}rnberg, D-91058 Erlangen, Germany} \affiliation{Physik-department, Technische Universit{\"a}t M{\"u}nchen, D-85748 Garching, Germany} \affiliation{D{\'e}partement de physique nucl{\'e}aire et corpusculaire, Universit{\'e} de Gen{\`e}ve, CH-1211 Gen{\`e}ve, Switzerland} \affiliation{Dept. of Physics and Astronomy, University of Gent, B-9000 Gent, Belgium} \affiliation{Dept. of Physics and Astronomy, University of California, Irvine, CA 92697, USA} \affiliation{Karlsruhe Institute of Technology, Institut f{\"u}r Kernphysik, D-76021 Karlsruhe, Germany} \affiliation{Dept. of Physics and Astronomy, University of Kansas, Lawrence, KS 66045, USA} \affiliation{SNOLAB, 1039 Regional Road 24, Creighton Mine 9, Lively, ON, Canada P3Y 1N2} \affiliation{Department of Physics and Astronomy, UCLA, Los Angeles, CA 90095, USA} \affiliation{Department of Physics, Mercer University, Macon, GA 31207-0001, USA} \affiliation{Dept. of Astronomy, University of Wisconsin, Madison, WI 53706, USA} \affiliation{Dept. of Physics and Wisconsin IceCube Particle Astrophysics Center, University of Wisconsin, Madison, WI 53706, USA} \affiliation{Institute of Physics, University of Mainz, Staudinger Weg 7, D-55099 Mainz, Germany} \affiliation{Department of Physics, Marquette University, Milwaukee, WI, 53201, USA} \affiliation{Institut f{\"u}r Kernphysik, Westf{\"a}lische Wilhelms-Universit{\"a}t M{\"u}nster, D-48149 M{\"u}nster, Germany} \affiliation{Bartol Research Institute and Dept. of Physics and Astronomy, University of Delaware, Newark, DE 19716, USA} \affiliation{Dept. of Physics, Yale University, New Haven, CT 06520, USA} \affiliation{Dept. of Physics, University of Oxford, Parks Road, Oxford OX1 3PU, UK} \affiliation{Dept. of Physics, Drexel University, 3141 Chestnut Street, Philadelphia, PA 19104, USA} \affiliation{Physics Department, South Dakota School of Mines and Technology, Rapid City, SD 57701, USA} \affiliation{Dept. of Physics, University of Wisconsin, River Falls, WI 54022, USA} \affiliation{Dept. of Physics and Astronomy, University of Rochester, Rochester, NY 14627, USA} \affiliation{Oskar Klein Centre and Dept. of Physics, Stockholm University, SE-10691 Stockholm, Sweden} \affiliation{Dept. of Physics and Astronomy, Stony Brook University, Stony Brook, NY 11794-3800, USA} \affiliation{Dept. of Physics, Sungkyunkwan University, Suwon 16419, Korea} \affiliation{Institute of Basic Science, Sungkyunkwan University, Suwon 16419, Korea} \affiliation{Dept. of Physics and Astronomy, University of Alabama, Tuscaloosa, AL 35487, USA} \affiliation{Dept. of Astronomy and Astrophysics, Pennsylvania State University, University Park, PA 16802, USA} \affiliation{Dept. of Physics, Pennsylvania State University, University Park, PA 16802, USA} \affiliation{Dept. of Physics and Astronomy, Uppsala University, Box 516, S-75120 Uppsala, Sweden} \affiliation{Dept. of Physics, University of Wuppertal, D-42119 Wuppertal, Germany} \affiliation{DESY, D-15738 Zeuthen, Germany}  \author{M. G. Aartsen} \affiliation{Dept. of Physics and Astronomy, University of Canterbury, Private Bag 4800, Christchurch, New Zealand} \author{M. Ackermann} \affiliation{DESY, D-15738 Zeuthen, Germany} \author{J. Adams} \affiliation{Dept. of Physics and Astronomy, University of Canterbury, Private Bag 4800, Christchurch, New Zealand} \author{J. A. Aguilar} \affiliation{Universit{\'e} Libre de Bruxelles, Science Faculty CP230, B-1050 Brussels, Belgium} \author{M. Ahlers} \affiliation{Niels Bohr Institute, University of Copenhagen, DK-2100 Copenhagen, Denmark} \author{M. Ahrens} \affiliation{Oskar Klein Centre and Dept. of Physics, Stockholm University, SE-10691 Stockholm, Sweden} \author{C. Alispach} \affiliation{D{\'e}partement de physique nucl{\'e}aire et corpusculaire, Universit{\'e} de Gen{\`e}ve, CH-1211 Gen{\`e}ve, Switzerland} \author{K. Andeen} \affiliation{Department of Physics, Marquette University, Milwaukee, WI, 53201, USA} \author{T. Anderson} \affiliation{Dept. of Physics, Pennsylvania State University, University Park, PA 16802, USA} \author{I. Ansseau} \affiliation{Universit{\'e} Libre de Bruxelles, Science Faculty CP230, B-1050 Brussels, Belgium}
% ...
\collaboration{IceCube Collaboration} \noaffiliation  

\begin{abstract}
Pulsar wind nebulae (PWNe) are the main gamma-ray emitters in the Galactic plane. They are diffuse nebulae that emit nonthermal radiation. Pulsar winds, relativistic magnetized outflows from the central star, shocked in the ambient medium produce a multiwavelength emission from the radio through gamma rays. Although the leptonic scenario is able to explain most PWNe emission, a hadronic contribution cannot be excluded. A possible hadronic contribution to the high-energy gamma-ray emission inevitably leads to the production of neutrinos. Using 9.5 yr of all-sky IceCube data, we report results from a stacking analysis to search for neutrino emission from 35 PWNe that are high-energy gamma-ray emitters. In the absence of any significant correlation, we set upper limits on the total neutrino emission from those PWNe and constraints on hadronic spectral components.\\
\end{abstract}


\section{Introduction} \label{sec:intro}
Galactic cosmic rays (CRs) are believed to reach energies of at least several PeV, the "knee" in the CR spectrum. Their interactions should generate gamma rays and neutrinos from the decay of secondary pions reaching hundreds of TeV. Because high-energy gamma rays can also originate in leptonic scenarios, the smoking gun for the identification of a Galactic cosmic accelerator relies on identifying a high-energy neutrino source.

The observation of high-energy neutrinos of astrophysical origin with IceCube \citep{Aartsen2013e,Aartsen:2014gkd} opened a new front in the search for Galactic CR accelerators. Since the discovery, IceCube has conducted analyses searching for the sources of cosmic neutrinos. 
Potential sources in the Galaxy are pre-identified from the catalogs of high-energy gamma-ray emitters in the Galaxy as high-energy gamma rays are supposed to be accompanied by their neutrino counterparts, if the sources are hadronic. In an early survey of the very-high-energy (VHE) gamma rays (100 GeV - 100 TeV) sky by Milagro \citep{Abdo:2006fq}, a handful of sources were identified as the brightest objects after the Crab Nebula. Early predictions hinted at the possibility of identifying these sources within a few years of IceCube operation \citep{Halzen:2008zj, GonzalezGarcia:2009jc}. Further observations by Milagro, together with other imaging atmospheric Cerenkov telescopes (IACTs) and water Cerenkov telescopes, surveys by the High Energy Stereoscopic System \citep[H.E.S.S.;][]{H.E.S.S.:2018zkf} and the High Altitude Water Cherenkov \citep[HAWC;][]{Abeysekara:2017hyn} for instance, provided a better and more comprehensive view of the Galactic plane at high energies. Interestingly, the majority of these objects were found to be pulsar wind nebulae (PWNe).

PWNe are diffuse nebulae confined inside supernova remnants (SNR) that are powered by pulsar winds generated by the highly spinning and magnetized pulsars in the center. According to observations mentioned above, PWNe are the most numerous TeV gamma-ray  emitters in the Milky Way.

 The photon emission of PWNe is believed to be mainly from relativistic electron-positron pairs, which are the primary components of the pulsar winds. These magnetized winds are powered by the rotational energy of the central pulsars. In this leptonic scenario, the low-energy emission (radio, optical, and X-ray) is dominated by synchrotron emission of relativistic leptons, and the inverse Compton scattering (ICS) of synchrotron photons becomes dominant at high energies (TeV). The leptonic scenario can accommodate the photon spectrum from radio wavelengths to TeV \citep{Kargaltsev:2010jy}. However, the presence of hadrons coexisting with leptons is still uncertain and to date cannot be excluded by either theory or observation. The hadronic mechanism was first discussed in the context of the VHE gamma-ray emission of the Crab Nebula,
 where protons accelerated in the outer gap of the pulsar interacting with the nebula \citep{Cheng:1990au} and heavy nuclei accelerated in the pulsar magnetosphere interacting with soft photons \citep{Bednarek:1997cn}. In addition, neutrino emission from PWNe has been studied for CR acceleration at the termination shocks followed by interactions in the source region with photons or nuclei \citep[see, e.g.,][]{ Guetta:2002hv,Amato:2003kw,Bednarek:2003cv,Lemoine:2014ala,DiPalma:2016yfy}.  Even minor contamination of ions at the termination shock would lead to considerable amount of energy contents released in hadrons. In such scenario, a neutrino flux is expected due to hadronuclear interactions \citep[see ][for details]{Amato:2006ts}.
    
VHE gamma-ray emission from PWNe and the possibility of their hadronic origin render PWNe of interest to IceCube. Previous IceCube searches have set upper limits on the neutrino flux from a list of individual PWNe \citep{Aartsen:2013uuv, Aartsen:2014cva,Aartsen:2016oji}  and also a stacking search on nine PWNe using 7 yr of data has been performed \citep{Aartsen:2017ujz}. Assuming part of the TeV gamma-ray emission from PWNe is hadronic, we report a stacking analysis on 35 TeV PWNe using 9.5 yr of IceCube all-sky neutrino data.

\section{SEARCH FOR NEUTRINO EMISSION}
\subsection{Source Selection}
In astrophysical beam dumps, when accelerated CRs interact with matter or ambient radiation, both neutral and charged pion secondaries are produced. While charged pions decay into high-energy neutrinos, neutral pions decay and create a flux of high-energy gamma rays. Therefore, in the context of multimessenger connection, high-energy neutrinos are inevitably accompanied by pionic gamma rays. PWNe with detected VHE gamma rays are of interest in the context of multimessenger astronomy, for possible hadronic origin in addition to photons scattered to higher energies via ICS. Therefore, in this search, we consider sources identified as PWNe with gamma-ray emission higher than 1 TeV. These are sources observed by the high-energy gamma-ray telescopes HAWC, H.E.S.S., Major Atmospheric Gamma-Ray Imaging Cherenkov (MAGIC), and Very Energetic Radiation Imaging Telescope Array System (VERITAS), which currently observe the highest energy photons. The associated pulsars of these PWNe are listed in the Australia Telescope National Facility (ATNF) catalog \citep{Manchester:2004bp}. The source list is presented in Table \ref{tab:sources} along with detailed information on the position, extension, age, period, and gamma-ray spectrum of each source. 

\subsection{Method}\label{sec:method}
Here, we use an unbinned maximum likelihood to perform a stacking search for neutrino emission from TeV PWNe. This analysis seeks a significant excess of neutrino events (signal) from directions interested above the background of atmospheric neutrinos and diffuse astrophysical neutrinos. 
The method is described in \cite{Braun:2008bg}. Stacking potential sources together is an effective way to improve the sensitivity of a search for neutrino sources \citep{Achterberg:2006ik}. The unbinned likelihood function for a stacking search is defined as
\begin{equation}
\label{eqn:llhfunc}
\mathcal{L}(n_{s}, \gamma_{s}) = \prod^{N}_{i}\left(\frac{n_{s}}{N}\sum^{M}_{j}\omega_{j}S_{i}^{j}+(1-\frac{n_{s}}{N})B_{i}\right),
\end{equation}
where $n_{s}$ is the number of signal events and $\gamma_{s}$ is the spectral index of a power-law spectrum. $N$ is the total number of neutrino events and $M$ is the number of sources. $S_{i}^{j}$ is the signal probability density function (PDF), which corresponds to the $i$th event with respect to the $j$th source. The normalized weight, $\omega_{j}$, determines the relative normalization of the signal PDF from source $j$. Finally, $B_{i}$ is the background PDF.  

The PDFs are composed of the spatial part and the energy part, therefore, for the signal $S_{i}^{j}=S^{s}(x_{j}, x_{i}, \sigma_{ij})\times S^{E}(E_{i}, \gamma_{s})$ and similarly for the background $B_{i}=1/(2\pi)B^{\delta}(\delta_{i})\times B^{E}(E_{i})$. $x_j$ is the location of source $j$; $x_i$, $\delta_i$ and $E_i$ are the reconstructed location, declination and energy of event $i$. For $S_{i}^{j}$, the spatial clustering of signal events is modeled as a two-dimensional Gaussian distribution. The width of the spatial PDF, $\sigma_{ij}$, representing the effective angular uncertainty of event $\sigma_i$ and the angular extension of source $\sigma_j$, is defined as $\sigma_{ij}=(\sigma_{i}^2+\sigma_{j}^2)^{1/2}$. An event energy proxy is used to separate a potential hard-spectrum signal from the softer spectrum background. We model the signal spectrum as an unbroken power-law spectrum, $E^{-\gamma_{s}}$, where the spectral index, $\gamma_{s}$, is assumed to have a value between 1 and 4. In order to avoid bias, we set the spectral index $\gamma_s$ as a generic parameter for all sources instead of using the measured index for each source from gamma-ray observations. For $B_i$, it is constructed from binning the experimental data in the reconstructed declination and energy. $1/2\pi$ arises due to IceCube, located at the South Pole, has a uniform acceptance in right ascension. Since we search for an excess of neutrino events from preassigned source locations, the background is estimated by randomizing the right ascensions of the experimental data sample to remove any correlation with sources being tested. The likelihood is maximized for two parameters: number of signal events, $n_{s}$, and the spectral index, $\gamma_s$. The null hypothesis presumes no signal-like event, i.e., $n_{s}=0$. The test statistic (TS) is defined by a maximal log-likelihood ratio, ${\rm TS}=2\log[\mathcal{L}(\hat{n}_{s},\hat{\gamma}_{s})/\mathcal{L}(n_{s}=0)]$ in which $\hat{n}_{s}$ and $\hat{\gamma}_{s}$ are the best-fitting values. A distribution of background TS values approximately following $\chi^2$ distribution can be generated after randomizing the neutrino map many times. The actual data can give the observed TS. The $p$-value, which represents the probability that the background being able to create a TS the same or larger than the observed TS, is defined as the fraction of TS larger than the observed one in the total background TS distribution. 

\subsection{Weighting}\label{sec:weighting}
The weight term $\omega_{j}$ is composed of two terms—a 
"model" term $\omega_{j,\mathrm{model}}$ and a detector acceptance term $\omega_{j,\mathrm{det}}$, i.e.,
\begin{equation}
\omega_{j}=\frac{\omega_{j,\mathrm{model}}\cdot\omega_{j,\mathrm{det}}}{\sum^{M}_{j}\omega_{j,\mathrm{model}}\cdot\omega_{j,\mathrm{det}}}.    
\end{equation}

The detector acceptance term $\omega_{j,\mathrm{det}}$ can be determined by the spectrum and the effective area of the detector for an event from the direction of the source—$\omega_{j,\mathrm{det}}\propto\int_{E_\mathrm{min}}^{E_\mathrm{max}}E^{-\gamma_{s}}A_{\mathrm{eff}}(\theta_j, E)dE$, where $\theta_j$ is the zenith angle of source $j$. About the unknown model term, $\omega_{j, \mathrm{model}}$, theoretical or observational arguments can be used in the weights applied to each source in order to test a specific hypothesis, such as correlation of neutrino emission with a particular property of a source. 
In a generic astrophysical beam dump, the production rate of neutrinos depends on the matter density and injection power of accelerated CRs. Furthermore, the capability of an accelerator to reach very high energies depends on how strongly it can confine the particles, as stated by the Hillas criterion \citep{hillas1984origin}. In the following we will adopt the main characteristics of PWNe that could attribute to these criteria and will test four distinct hypotheses by incorporating different weighting schemes according to these properties:\\
  
\begin{figure}[t!]
\centering
\includegraphics[width=1.0\columnwidth]{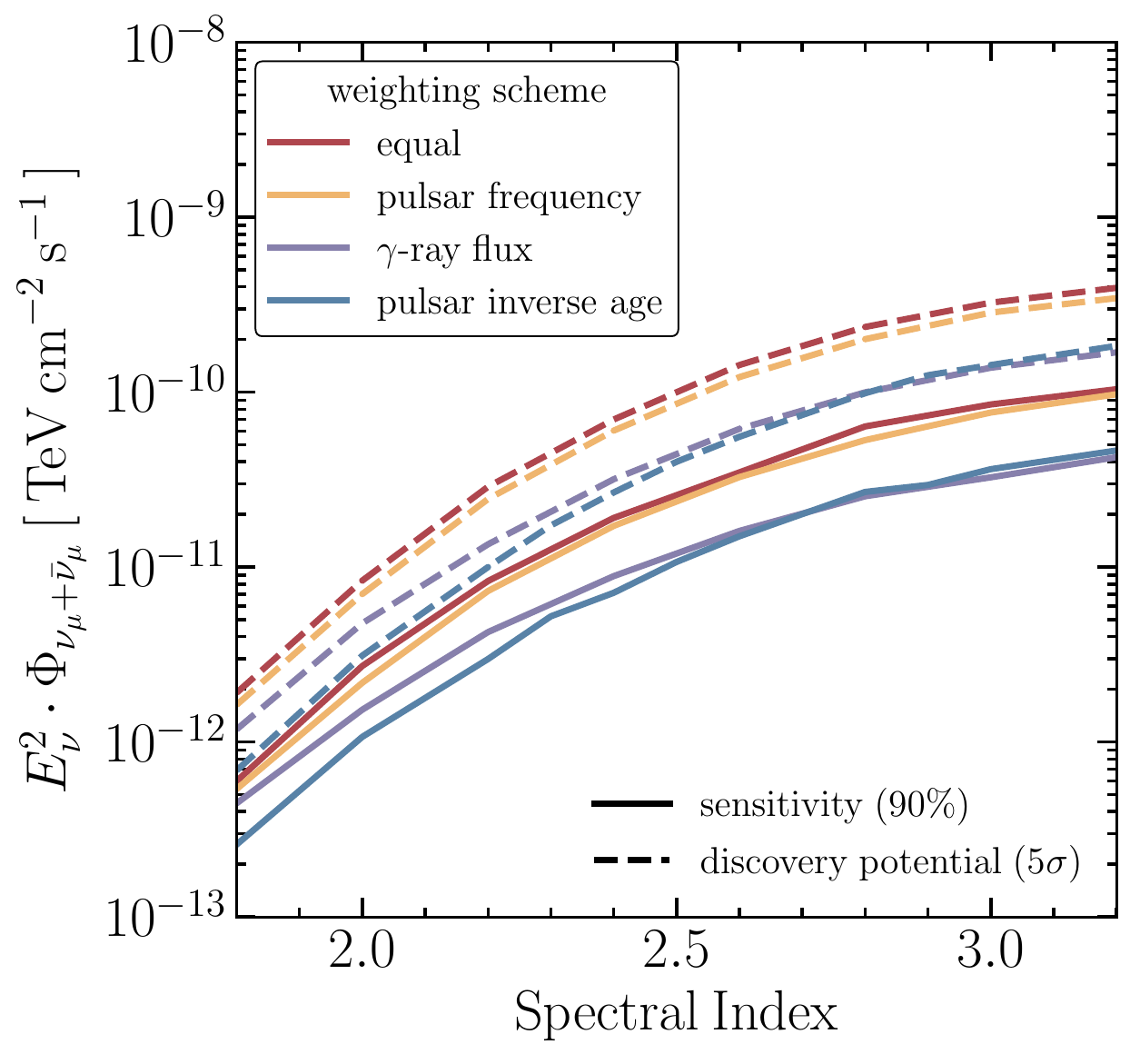}
\caption{Sensitivities (90$\%$ CL) and 5$\sigma$ discovery potentials of different weighting schemes as a function of spectral index for an unbroken power-law spectrum. \label{fig:sensitivity}}
\end{figure}

{\bf{Equal weighting—}}  
In this scheme, $\omega_{j, \mathrm{model}}=1$, which assigns the neutrino emission of each source the same probability. Therefore, no preference is given to any source and they are treated equally.

{\bf{Gamma-ray flux weighting—}}  
This case assumes that a plausible high-energy neutrino emission is directly proportional to the high-energy gamma-ray emission from each source. If this assumption is true, this basically means that the observed high-energy gamma rays are either partially or completely of hadronic origin. Here, we incorporate the gamma-ray flux at 1 TeV as the weights. As indicated in Table \ref{tab:sources}, for sources in the Northern sky, spectral measurements from MAGIC and VERITAS are used; while for sources in the Southern sky, H.E.S.S. is used as it is more sensitive in this region. The HAWC observations are used for Geminga and 2HWC J0700+143. 

{\bf{Pulsar spin frequency weighting-}}
 The energy carried by the pulsar wind for the acceleration is taken from the rotational energy of the pulsar as it emits radiation, which results in the spin-down of the pulsar \citep{Gaensler:2006ua}. Faster spinning pulsars are more energetic, and are considered as candidate sources of ultra-high-energy CRs, see, for example \cite{Bednarek:2001av,Kotera:2015pya}. In addition, the acceleration time has a strong dependence on the period. Therefore, the period of the pulsar, as an important measure of how energetic the pulsar is, is used as the weight in this scheme. In this scheme, faster spinning sources are preferred.

{\bf{Age weighting-}}
 The characteristic age of a pulsar is usually defined as $\tau=\frac{P}{2\dot{P}}$ where $P$ and $\dot{P}$ are the period and its time derivative, respectively. This parameter is used to estimate the true age of a pulsar under assumptions that the initial spin is much faster than today and the energy loss is from magnetic dipole radiation 
 \citep{Gaensler:2006ua}. Following the discussion of spin frequency weighting scheme, the age of a pulsar presents another factor that may determine how energetic it is, as the age is not only dependent on the period but also on its time derivative. Given that young fast-spinning neutron star winds have been proposed as the sites for CR acceleration \citep{Blasi:2000xm, Bednarek:2003cv, Fang:2015xhg}. Here, we use 1/age as the weight for each source. This assumption prefers younger PWNe to be more energetic emitters. This hypothesis is in accordance with the idea that young and highly magnetized pulsars are primary sources of ultra-high-energy CRS. On the other hand, this scenario may also examine the hypothesis that 
accelerated particles are not currently part of the pulsar wind, instead, they have been injected in the nebula at some earlier time when the pulsar was much younger and more energetic \citep{1999A&A...346L..49A}.

\subsection{Detector \& Data Set}

\begin{deluxetable}{lDr}[b]
\tablecaption{IceCube Data Set \label{tab:icperiods}}
\tablewidth{0pt}
\tablehead{
\colhead{Sample} & \multicolumn2c{Livetime} & \colhead{Events}\\ 
\colhead{}&\multicolumn2c{days} & \colhead{\#}
}
\decimals
\startdata
IC 40 & 376.36 & 36900\\
IC 59 & 353.58 & 107011\\
IC 79 & 316.05 & 93133\\
IC 86 \uppercase\expandafter{\romannumeral 1} & 332.96  & 136244\\
IC 86 \uppercase\expandafter{\romannumeral 2} & 1058.48 & 338590\\
GFU 2015-2017 & 989.95 & 571040\\
\enddata
\tablecomments{The data set used in this search. The first seven years of data, IC40-IC86 II, are the same as data used in \citep{Aartsen:2016oji,Aartsen:2017ujz} and the latter 2.5 years, GFU 2015-2017, are discussed in \citep{Aartsen:2016lmt}.}
\end{deluxetable}

\begin{deluxetable*}{lDDccDDD}[t]
\tablecaption{Results}
\tablewidth{9pt}
\tablehead{
\colhead{weighting} & \multicolumn2c{TS} & \multicolumn2c{$n_{s}$} & \colhead{$\gamma$} & \colhead{$p$-value} & \multicolumn2c{$\Phi_{\nu_{\mu}+\bar{\nu}_{\mu}}^{90\%,\;E^{-2.0}}$} & \multicolumn2c{$\Phi_{\nu_{\mu}+\bar{\nu}_{\mu}}^{90\%,\;E^{-2.19}}$} & \multicolumn2c{$\Phi_{\nu_{\mu}+\bar{\nu}_{\mu}}^{90\%,\;E^{-2.5}}$}
}
\decimals
\startdata
Equal & 0.81 & 40.43 & 3.84 & 23\% & 3.91 & 11.6 & 44.5 \\
Frequency & 0.26 & 18.00 & 3.81 & 38\% & 2.64 & 7.79 & 28.2 \\
Flux & 0.21 & 8.73 & 4.00 & 36\% & 1.74 & 4.57 & 14.9 \\
Inverse age & 0 & 0 & - & - & 1.07 & 2.82 & 10.7 \\
\enddata
\tablecomments{Best fits for TS, $n_{s}$ and $\gamma$. The last three columns are upper limit constraints on the stacking flux with a 90\% CL. The first one has a power-law spectrum $E^{-2.0}$; the second has $E^{-2.19}$, which is the measured astrophysical muon neutrino spectrum by IceCube \citep{Haack:2017dxi} and the last column follows $E^{-2.5}$, which is the IceCube all-flavor combined neutrino spectrum \citep{Aartsen:2015knd}. They are all normalized at 1$\;\rm{TeV}$ with units $10^{-12}\;\rm{TeV^{-1}cm^{-2}s^{-1}}$.} 
\label{tab:Results}
\end{deluxetable*}

The IceCube Neutrino Observatory at the South Pole has transformed a cubic kilometer of Antarctic ice into a Cerenkov detector and has been monitoring the whole sky continuously since 2008. The detector is an array of digital optical modules (DOMs) each including a photomultiplier tube and on-board read-out electronics \citep{Abbasi:2010vc, Abbasi:2008aa}. The complete configuration accomplished in 2010 is composed of 5160 DOMs arranged in 86 strings from 1450 - 2450 m below the surface in Antarctic ice \citep{Aartsen:2016nxy}. The Cerenkov light emitted by the secondary particles produced in neutrino interactions are registered by the DOMs, and particle trajectories are determined by the arrival times of photons at the optical sensors. The number of photons observed along with their timings are used to determine the energy deposited by charged secondary particles in the detector. While IceCube is able to detect neutrinos of all flavors, long tracks resulting from muon neutrino interactions can point back to the sources with a typical angular resolution of less than 1$^\circ$ \citep{Aartsen:2016oji}.

In this analysis, we use 9.5 yr all-sky data collected by IceCube between 2008 April and 2017 November. This includes 7 yr of data already studied for neutrino point sources \citep{Aartsen:2016oji} along with additional data for the period from 2015 May to 2017 November \citep{Aartsen:2016lmt}. These 9.5 yr of data correspond to six distinct periods specified in Table \ref{tab:icperiods}. These periods differ in detector configuration, data-taking conditions, and event selections. 

To estimate the performance of the analysis, source emission is simulated to observe the detector response. Sensitivities (90\% confidence level (CL)) and discovery potentials (5$\sigma$) for different weighting scenarios discussed in Sec.~\ref{sec:weighting} are shown in Figure \ref{fig:sensitivity}. 
For simulating the neutrino emission, an unbroken power-law spectral shape is assumed. The projected sensitivity shows, as expected, that IceCube is more sensitive to sources with a harder spectrum. The difference of the sensitivities between weighting schemes is dependent on the weight distribution, which represents how significant we assume one location is. For example, more sources in the Northern sky with higher weights imply a better sensitivity to IceCube.  

\section{RESULTS}
We performed the unbinned likelihood analysis discussed in Sec.~\ref{sec:method} for different hypotheses of neutrino emission considering equal, frequency, gamma-ray flux, and inverse age weighting. The results for these tests are presented in Table \ref{tab:Results}. The largest excess was found in the equal weighting scheme, yielding a fitted signal of 40.4 events with a $p$-value of 0.22, which shows no significant correlation. Therefore, the isotropic background distribution (null) hypothesis is preferred. Because none of the tests led to a significant excess of fitted signal events and the results are compatible with the null hypothesis, we set upper limits on the total flux of high-energy neutrinos from PWNe for each hypothesis. Upper limits with a 90\% CL are presented for three different spectral shape assumptions in Table \ref{tab:Results}.\\

\section{DISCUSSION}
\begin{figure}[ht!]
\centering
\includegraphics[width=1.0\columnwidth]{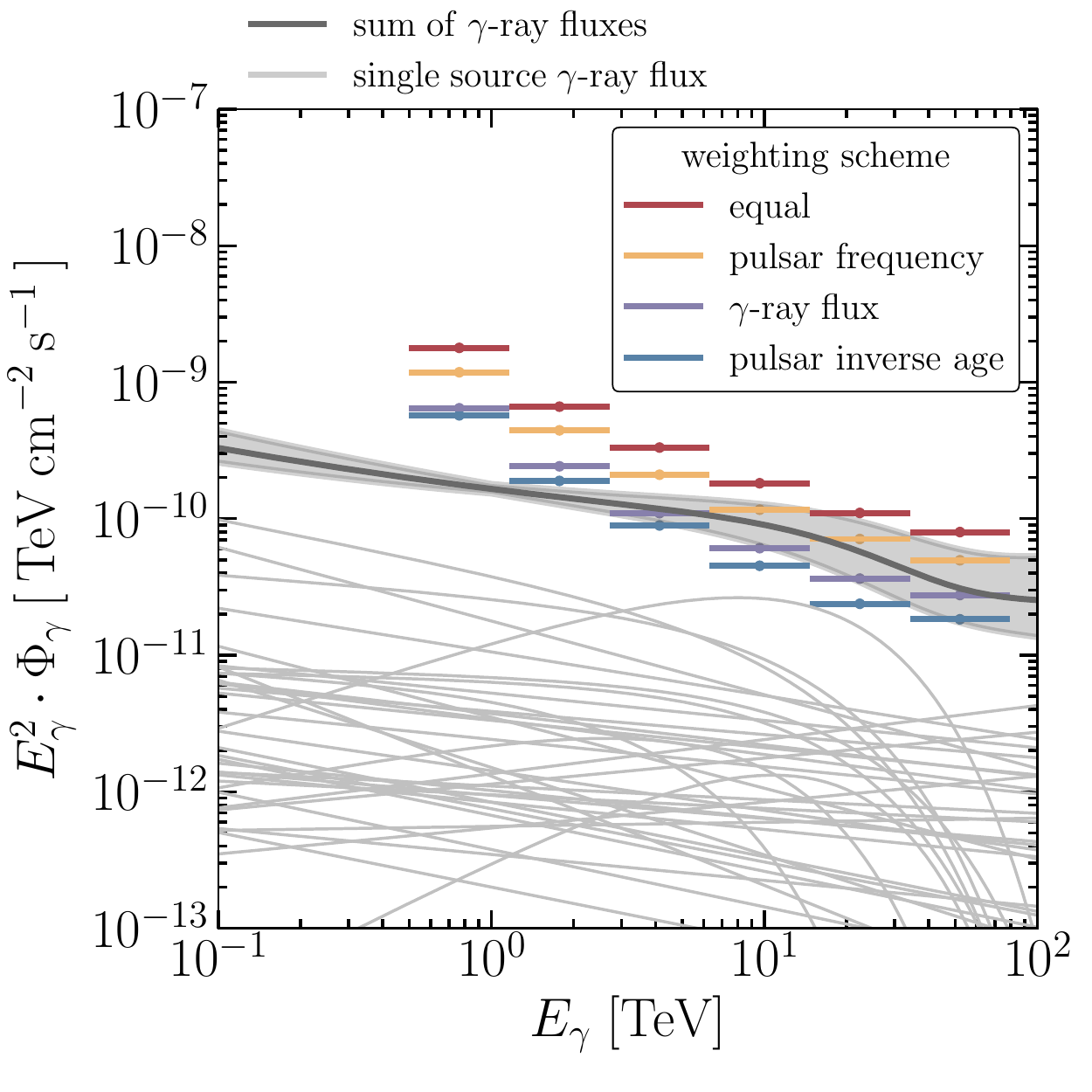}
\caption{ Light gray lines are observed gamma-ray spectra of the sources in this search, and the dark gray line is the sum of those fluxes. The total uncertainty is estimated by simply summing up the uncertainty of the flux of each source. Red, orange, purple, and blue steps show the differential upper limit on the hadronic gamma-ray emission. The upper limits are obtained by converting 90\% CL differential upper limit on the neutrino flux, and each color corresponds to a given weighting method. To avoid uncertainties from extrapolation, the energy is limited to 100 TeV here.} 
 \label{fig:edifful}
\end{figure}
Due to the apparent isotropy of astrophysical neutrinos observed by IceCube, An extragalactic origin is expected to be predominant. However, Galactic CR accelerators are expected to contribute at a subdominant level to the observed high-energy cosmic neutrino flux. The Galactic component of the high-energy neutrino flux is constrained to $\sim14\%$ at 1 TeV  \citep{Aartsen:2017ujz} of the combined diffuse neutrino flux measured in \cite{Aartsen:2015knd}. The upper limit obtained in this study for neutrino emission from TeV PWNe is consistent with this limit by showing no more than $\sim1.6\%$ contribution to the combined flux in this search. Considering the astrophysical muon neutrino flux reported \citep{Haack:2017dxi}, the contribution of neutrino emission from TeV PWNe studied here to the total neutrino flux is less than $\sim 4\%$. We note that this limit is valid within the specific assumptions of this analysis regarding the weighting and selection of the sources and should not be applied or extended to other hypotheses.

In the context of a multimessenger connection, neutrino fluxes can be related to the high-energy gamma-ray flux; see \cite{Ahlers:2013xia} for details. With this connection, one can use the upper limit on the neutrino flux to constrain the hadronic component of the observed high-energy gamma-ray flux. Here, we assume hadronuclear (i.e., proton–proton) interactions at the source to convert neutrino fluxes to their gamma-ray counterparts. For proton–gamma interactions, one can easily adjust proton–proton flux by a factor of 2, taking account the different ratio of charged to neutral pions in each process.   

 High-energy gamma-ray flux measurements extend to tens of TeV, while IceCube neutrinos reach energies of a few PeV. To avoid the large uncertainties in extrapolation of the high-energy gamma-ray flux, we calculate differential upper limits assuming an unbroken power-law spectrum and convert the neutrino limits into upper limits on a hadronic gamma-ray flux at energies below 100 TeV. Figure \ref{fig:edifful} shows the differential upper limits for an $E^{-2}$ spectrum for different hypotheses tests of this study compared to the observed cumulative flux of VHE gamma rays.  As expected, the constraints are stronger at higher energies. At energies between 10 and 100 TeV, the hadronic component of the high-energy emission from these sources are constrained, if the neutrino emission is either correlated with the observed gamma-ray emission or if younger PWNe are more efficient neutrino emitters. However, if the emission is proportional to the pulsar's frequency, upper limits are marginally at the same level of the total gamma-ray emission.  
 \\

\section{Conclusion \& Outlook}

Galactic CRs reach energies of at least several PeV, and their interactions should generate gamma rays and neutrinos from the decay of secondary pions. Therefore, Galactic sources are expected to contribute at some level to the total high-energy cosmic neutrino flux observed by IceCube. In the initial survey of the VHE sky by Milagro \citep{Abdo:2006fq}, where the observed gamma-ray flux in TeV was found higher than the expected leptonic emission, promising sources had been identified based on their spectra, assuming that the highest energy gamma rays are pionic. Early estimates showed that the observation of these sources were likely in the lifetime of IceCube \citep{Halzen:2008zj,GonzalezGarcia:2009jc}. Further observation of the Galactic plane by IACTs provided  more insight, and updated estimates showed that IceCube would identify those sources provided that the gamma-ray fluxes did not cut off at low energies \citep{Gonzalez-Garcia:2013iha, Halzen:2016seh}. Meanwhile, the majority of the sources in the plane were identified to be PWNe. Leptonic scenarios are generally more favored for describing the high-energy emission from PWNe. However, a hadronic component cannot be excluded by current observations. Hadronic interactions at the source will inevitably result in the production of neutrinos that provide the smoking gun for the presence of the hadrons.

In this study, we examined the possible neutrino emission from PWNe with TeV gamma-ray emission. Thirty-five sources were identified, and the results of the stacking searches for the high-energy neutrino emission are compatible with the isotropic arrival direction hypothesis. In the absence of a significant excess of neutrino events in the direction of these sources, we have set upper limits on the total neutrino emission and on the potential hadronic component of the high-energy gamma-ray flux.

 Any evidence for presence (or absence) of the hadrons in pulsar winds would provide important clues about the mechanism of acceleration in these sources, for more details see, e.g. \citep{Amato:2013fua}. The so-called $\sigma$ problem \footnote{$\sigma$ problem refers to the conflicting scenario in theoretical modelings of pulsar wind. $\sigma$ presents the ratio of the wind Poynting flux to its kinetic energy flux. While theoretical models of pulsar magnetospheres and wind predict large $\sigma$, the 1D MHD simulations of PWN cannot match shock size and expansion speed at same time with high sigma.} could be solved if the majority of the pulsar winds energy is carried by hadrons and further explain how efficient acceleration of leptons is obtained in the termination shocks \citep{DiPalma:2016yfy}. The stacking analysis presented here found upper limits at the level of the total observed high-energy gamma-ray emission indicating that neutrino flux measurements getting close to determine the feasibility of such models.

 In the future, more accurate measurement of the VHE gamma-ray flux by HAWC and coming gamma-ray observatories, such as CTA \citep{Acharya:2017ttl} and LHAASO \citep{DiSciascio:2016rgi}, will shed more light on the nature of the high-energy emission from the Milky Way. From the perspective of neutrino detection, the addition of more years of data with continuous operation of IceCube will improve the sensitivity of the search for Galactic sources of cosmic neutrinos. The next step, IceCube-Gen2, a substantial expansion of IceCube, will be 10 times larger. This next-generation neutrino observatory with five times the effective area of IceCube is expected to improve the neutrino source search sensitivity by the same order \citep{Aartsen:2014njl, Ahlers:2014ioa, Aartsen:2019swn}. With higher neutrino statistics, identifying Galactic sources will become more promising. \\

\section*{Acknowledgements }
The IceCube collaboration acknowledges the significant contributions to this manuscript from Ali Kheirandish and Qinrui Liu. The authors gratefully acknowledge the support from the following agencies and institutions: USA {\textendash} U.S. National Science Foundation-Office of Polar Programs, U.S. National Science Foundation-Physics Division, Wisconsin Alumni Research Foundation, Center for High Throughput Computing (CHTC) at the University of Wisconsin-Madison, Open Science Grid (OSG), Extreme Science and Engineering Discovery Environment (XSEDE), U.S. Department of Energy-National Energy Research Scientific Computing Center, Particle astrophysics research computing center at the University of Maryland, Institute for Cyber-Enabled Research at Michigan State University, and Astroparticle physics computational facility at Marquette University; Belgium {\textendash} Funds for Scientific Research (FRS-FNRS and FWO), FWO Odysseus and Big Science programmes, and Belgian Federal Science Policy Office (Belspo); Germany {\textendash} Bundesministerium f{\"u}r Bildung und Forschung (BMBF), Deutsche Forschungsgemeinschaft (DFG), Helmholtz Alliance for Astroparticle Physics (HAP), Initiative and Networking Fund of the Helmholtz Association, Deutsches Elektronen Synchrotron (DESY), and High Performance Computing cluster of the RWTH Aachen; Sweden {\textendash} Swedish Research Council, Swedish Polar Research Secretariat, Swedish National Infrastructure for Computing (SNIC), and Knut and Alice Wallenberg Foundation; Australia {\textendash} Australian Research Council; Canada {\textendash} Natural Sciences and Engineering Research Council of Canada, Calcul Qu{\'e}bec, Compute Ontario, Canada Foundation for Innovation, WestGrid, and Compute Canada; Denmark {\textendash} Villum Fonden, Danish National Research Foundation (DNRF), Carlsberg Foundation; New Zealand {\textendash} Marsden Fund; Japan {\textendash} Japan Society for Promotion of Science (JSPS) and Institute for Global Prominent Research (IGPR) of Chiba University; Korea {\textendash} National Research Foundation of Korea (NRF); Switzerland {\textendash} Swiss National Science Foundation (SNSF); United Kingdom {\textendash} Department of Physics, University of Oxford. \\

\bibliography{bibfile}

\begin{longrotatetable}
\movetabledown = 1cm
\begin{deluxetable*}{ccDDDDDDDDcc}
\centering
\tablecaption{Sources in this search.\label{tab:sources}}
\tablewidth{9pt}
\tablehead{
\colhead{PWN} & \colhead{Pulsar} & 
\multicolumn2c{R.A.} & \multicolumn2c{Dec.} & 
\multicolumn2c{Extension} & \multicolumn2c{Period} & 
\multicolumn2c{Age} & \multicolumn2c{$N_{0}$} & 
\multicolumn2c{$\gamma$} & \multicolumn2c{Cutoff} & \colhead{Telescope}&\colhead{Ref.} \\ 
\colhead{} & \colhead{} & \multicolumn2c{$\deg$} & \multicolumn2c{$\deg$} & 
\multicolumn2c{$\deg$} & \multicolumn2c{s} & \multicolumn2c{kyr} &
\multicolumn2{c}{$\mathrm{TeV}^{-1}\mathrm{cm}^{-2}\mathrm{s}^{-1}$} & \multicolumn2c{} & \multicolumn2c{TeV} & \colhead{}&\colhead{}}
\decimals
\startdata
CTA1 & J0007+7303 & 1.61 & 72.98 & 0.25 & 0.316 & 13.9 & $\qquad$\centering 0.10 & 2.2 & - & VERITAS & \citep{Aliu:2012uj} \\
3C 58 & J0205+6449 & 31.38 & 64.85 & - & 0.0657 & 5.37 & \centering 0.02 & 2.4 & -  & MAGIC & \citep{Aleksic:2014pza}  \\
Crab & B0531+21 & 83.63 & 22.01 & - & 0.033 & 1.26 & 3.76 & 2.39 & 14.3 & HESS & \citep{Aharonian:2006pe} \\
LHA 120-N 157B & J0537-6910 & 84.43 & -69.17 & 0.014 & 0.0161 & 4.93 & 0.13 & 2.8 & -  & HESS & \citep{Abramowski:2015rca} \\
Geminga & J0633+1746  & 98.12 & 17.37 & 2.0 & 0.237 & 342.0 & 0.37 & 2.23 & -  & HAWC & \citep{Abeysekara:2017hyn}  \\
2HWC J0700+143 & B0656+14 & 105.12  & 14.32 & 1.0 & 0.385 & 111.0 & 0.094 & 2.17 & - &HAWC & \citep{Abeysekara:2017hyn}  \\
Vela X & B0833-45 & 128.75 & -45.6 & 0.59 & 0.089 & 11.3 & 1.21 & 1.35 & 12.27 & HESS & \citep{H.E.S.S.:2018zkf}  \\
HESS J1018-589B & J1016-5857 & 154.13  & -58.98 & 0.15 & 0.107 & 21.0 & 0.084 & 2.2 & - & HESS & \citep{H.E.S.S.:2018zkf}  \\
HESS J1026-582 & J1028-5819 & 156.66 & -58.2 & 0.13 & 0.0914 & 90.0 &  0.054 & 1.81 &  - & HESS & \citep{H.E.S.S.:2018zkf} \\
SNR G292.2-00.5 & J1119-6127  & 169.75 & -61.4 & 0.098 & 0.408 & 1.61 & 0.15 & 2.64 & -  & HESS & \citep{H.E.S.S.:2018zkf} \\
HESS J1303-631 & J1301-6305 & 195.7 & -63.18 & 0.177 & 0.185 & 11.0 & 0.63 & 2.04 & 15.12 & HESS &  \citep{H.E.S.S.:2018zkf} \\
HESS J1356-645 & J1357-6429 & 209.0 & -64.5 & 0.23 & 0.166 & 7.31 & 0.53 & 2.2 & - & HESS & \citep{H.E.S.S.:2018zkf}\\
Kookaburra(Rabbit) & J1418-6058 & 214.52 & -60.98  & 0.108 & 0.111 & 10.3 & 0.34 & 2.26 &  - & HESS & \citep{H.E.S.S.:2018zkf}  \\
Kookaburra(PWN) & J1420-6048 & 215.04 & -60.76 & 0.081 & 0.0682 & 13.0 & 0.33 & 2.2 & - & HESS & \citep{H.E.S.S.:2018zkf}  \\
HESS J1458-608 & J1459-6053 & 224.54 & -60.88 & 0.373 & 0.103 & 64.7 & 0.11 & 1.81 &  - & HESS &  \citep{H.E.S.S.:2018zkf} \\
MSH15-52 & B1509-58 &228.53 & -59.16 & 0.15 & 0.151 & 1.56 & 0.69 & 2.05 & 19.2 & HESS & \citep{H.E.S.S.:2018zkf}  \\
SNR G327.1-01.1\tablenotemark{a} & - & 238.65  & -55.08 & - & 0.035 & 18.0 & 0.035 & 2.19 &  - & HESS & \citep{H.E.S.S.:2018zkf}  \\
HESS J1616-508 & J1617-5055 & 244.1  & -50.9 & 0.232 & 0.0693 & 8.13 & 1.06 & 2.32 & -  & HESS & \citep{H.E.S.S.:2018zkf}  \\
HESS J1632-478 & J1632-4757 & 248.04 & -47.82 & 0.182 & 0.229 & 240.0 & 0.35 & 2.52 &  - & HESS & \citep{H.E.S.S.:2018zkf}  \\
HESS J1640-465 & J1640-4631 & 250.18 & -46.53 & 0.11 & 0.206 & 3.35 & 0.45 & 2.12 & 4.13 & HESS & \citep{H.E.S.S.:2018zkf}  \\
HESS J1708-443 & B1706-44  & 257.05 & -44.33 & 0.279 & 0.102 & 17.5 & 0.39 & 2.17 & - & HESS & \citep{H.E.S.S.:2018zkf} \\
HESS J1718-385 & J1718-3825 & 259.53 & -38.55 & 0.115 & 0.0747 & 89.5 & 0.030 & 0.98 & 10.57 & HESS &\citep{H.E.S.S.:2018zkf}  \\
SNR G000.9+00.1 & J1747-2809 & 266.85 & -28.15 & - & 0.0521 & 5.31 & 0.084 & 2.4 & - & HESS & \citep{H.E.S.S.:2018zkf}  \\
HESS J1813-178 & J1813-1749 & 273.4 & -17.84 & 0.049 & 0.0447 & 5.6 & 0.22 & 1.64 & 7.37 & HESS & \citep{H.E.S.S.:2018zkf}  \\
HESS J1825-137 & B1823-13 & 276.42 & -13.84 & 0.461 & 0.101 & 21.4 & 2.56 & 2.15 & 13.57 & HESS &\citep{H.E.S.S.:2018zkf}  \\
HESS J1831-098 & J1831-0952 & 277.85 & -9.9 & 0.15 & 0.0673 & 128.0 & 0.11 & 2.1 & -  & HESS &  \citep{Sheidaei:2011vg} \\
HESS J1833-105 & J1833-1034 & 278.39   & -10.56 & - & 0.0619 & 4.85 & 0.038 & 2.42 &  - & HESS & \citep{H.E.S.S.:2018zkf} \\
HESS J1837-069 & J1838-0655  & 279.41 & -6.95 & 0.355 & 0.0705 & 22.7 & 1.78 & 2.54 &  - & HESS & \citep{H.E.S.S.:2018zkf} \\
HESS J1846-029 & J1846-0258 & 281.6  & -2.98 & -  & 0.327 & 0.73 & 0.067 & 2.41 &  - & HESS & \citep{H.E.S.S.:2018zkf}  \\
IGR J18490-0000 & J1849-0001 & 282.24 & -0.04 & 0.09 & 0.0385 & 42.9 & 0.056 & 1.97 & - & HESS &\citep{H.E.S.S.:2018zkf}  \\
MAGICJ1857.2+0263 & J1856+0245 & 284.3  & 2.63 & 0.1 & 0.0809 & 20.6 & 0.24 & 2.2 & - & MAGIC &\citep{Aleksic:2014rza} \\
SNR G054.1+00.3 & J1930+1852 & 292.63 & 18.87 & -  & 0.137 & 2.89 & 0.075 & 2.39 & - & VERITAS &\citep{Acciari:2010qs} \\
MGRO J2019+37 & J2021+3651 & 304.65 & 36.83 & 0.75 & 0.104 & 17.2 & 0.14 & 1.75  & - & VERITAS &\citep{Aliu:2014rha} \\
TeV J2032+4130 & J2032+4127  & 308.03 & 41.51 & 0.158 & 0.143 & 201.0 & 0.095 & 2.1 &  - & VERITAS &\citep{Aliu:2014pta}  \\
Boomerang & J2229+6114 & 337.18 & 61.17 & 0.22 & 0.0516 & 10.5 & 0.14 & 2.29 &  - & VERITAS & \citep{Acciari:2009zz} \\
\enddata
\tablecomments{Equatorial coordinates (J2000) of sources are from TeVCat. Ages and periods that are related to the associated pulsars are from ATNF \citep{Manchester:2004bp}. The last two columns refer to telescopes and references that made spectrum (normalized fluxes, spectral indices, and exponential cutoff energies, if applicable) and angular extension measurements of the source. $N_0$ in the table shows the gamma-ray flux at 1 TeV.}
\tablenotetext{a}{Period and age are from estimation \citep{Acero:2012sh}.}
\end{deluxetable*}
\end{longrotatetable}

\end{document}